\PassOptionsToPackage{table}{xcolor}
\documentclass[manuscript]{acmart}

\AtBeginDocument{%
  \providecommand\BibTeX{{%
    \normalfont B\kern-0.5em{\scshape i\kern-0.25em b}\kern-0.8em\TeX}}}

\newcommand{\parbold}[1]{\vspace{2.7mm}\textbf{#1.}}

\setcopyright{rightsretained}

\begin{document}

\title{Predicting Music Relistening Behavior Using the ACT-R Framework}

\author{Markus Reiter-Haas}
\email{reiter-haas@tugraz.at}
\affiliation{%
  \institution{Graz University of Technology}
  \streetaddress{Rechbauerstraße 12}
  \city{Graz}
  \state{Styria}
  \country{Austria}
  \postcode{8010}
}

\author{Emilia Parada-Cabaleiro}
\email{emilia.parada-cabaleiro@jku.at}
\author{Markus Schedl}
\email{markus.schedl@jku.at}
\affiliation{%
  \institution{Johannes Kepler University Linz (JKU) and Linz Institute of Technology (LIT)}
  \streetaddress{Altenbergerstraße 69}
  \city{Linz}
  \state{Upper Austria}
  \country{Austria}
  \postcode{4040}
}

\author{Elham Motamedi}
\email{elham.motamedi@famnit.upr.si}
\author{Marko Tkalcic}
\email{marko.tkalcic@famnit.upr.si}
\affiliation{%
  \institution{University of Primorska}
  \streetaddress{Titov trg 4}
  \city{Koper}
  \country{Slovenia}
  \postcode{6000}
}

\author{Elisabeth Lex}
\email{elisabeth.lex@tugraz.at}
\affiliation{%
  \institution{Graz University of Technology}
  \streetaddress{Rechbauerstraße 12}
  \city{Graz}
  \state{Styria}
  \country{Austria}
  \postcode{8010}
}

\renewcommand{\shortauthors}{M. Reiter-Haas, E. Parada-Cabaleiro, M. Schedl, E. Motamedi, M. Tkalcic, and E. Lex}

\copyrightyear{2021} 
\acmYear{2021} 
\acmConference[RecSys '21]{Fifteenth ACM Conference on Recommender Systems}{September 27-October 1, 2021}{Amsterdam, Netherlands}
\acmBooktitle{Fifteenth ACM Conference on Recommender Systems (RecSys '21), September 27-October 1, 2021, Amsterdam, Netherlands}\acmDOI{10.1145/3460231.3478846}
\acmISBN{978-1-4503-8458-2/21/09}

\begin{abstract}
  Providing suitable recommendations is of vital importance to improve the user satisfaction of music recommender systems. Here, users often listen to the same track repeatedly and appreciate recommendations of the same song multiple times. Thus, accounting for users' relistening behavior is critical for music recommender systems. In this paper, we describe a psychology-informed approach to model and predict music relistening behavior that is inspired by studies in music psychology, which relate music preferences to human memory. We adopt a well-established psychological theory of human cognition that models the operations of human memory, i.e., Adaptive Control of Thought—Rational (ACT-R). In contrast to prior work, which uses only the base-level component of ACT-R, we utilize five components of ACT-R, i.e., base-level, spreading, partial matching, valuation, and noise, to investigate the effect of five factors on music relistening behavior: (i) recency and frequency of prior exposure to tracks, (ii) co-occurrence of tracks, (iii) the similarity between tracks, (iv) familiarity with tracks, and (v) randomness in behavior. On a dataset of 1.7 million listening events from Last.fm, we evaluate the performance of our approach by sequentially predicting the next track(s) in user sessions. We find that recency and frequency of prior exposure to tracks is an effective predictor of relistening behavior. Besides, considering the co-occurrence of tracks and familiarity with tracks further improves performance in terms of R-precision. We hope that our work inspires future research on the merits of considering cognitive aspects of memory retrieval to model and predict complex user behavior.
\end{abstract}



\maketitle

\section{Introduction and Research Context}

Music recommender systems (MRS)~\cite{schedl2014music} are nowadays employed in many use cases, such as automatic playlist generation, next-track recommendation, context-aware music recommendation, and even in the creative process of music production~\cite{Schedl2017}. Providing useful music recommendations is a challenging task due to various aspects such as the variability in the purpose of music consumption, inconclusive and insufficient user feedback, or situational and contextual aspects that influence a user's current music preference~\cite{DBLP:journals/ijmir/SchedlZCDE18,lex2020modeling}.
Besides, users of MRS show a tendency to \textit{relisten to songs}, which means they appreciate recommendations of the same song multiple times~\cite{tsukuda2020explainable,conrad2019extreme}, which is in stark contrast to movie or product recommendation systems, where users typically do not prefer repeated recommendations~\cite{DBLP:journals/ijmir/SchedlZCDE18,ren2019repeatnet}.

In this work, we introduce a novel psychology-informed approach to model and predict \textit{music relistening} behavior of tracks. 
Our work is inspired by related work on psychology-informed recommender systems~\cite{lex2021psychology}, and studies in music psychology, which find that music preferences are biased by human memory: user studies showed that repeated exposure increases recognition and positive attitude towards music~\cite{peretz1998exposure}. 

While several works have been proposed to model repeated consumption of items, such as temporal point processes~\cite{wang2019modeling}, factorization models~\cite{rafailidis2015repeat} or neural networks~\cite{ren2019repeatnet}, to the best of our knowledge, they do not incorporate underlying cognitive processes of human behavior into their approaches.

In our work, we exploit a theory about human cognition, i.e., the cognitive architecture \textit{Adaptive Control of Thought—Rational (ACT-R)}~\cite{anderson2004integrated}, and investigate the utility of its declarative memory module for the task at hand. The declarative memory module of ACT-R stores and retrieves information and consists of separate components. In contrast to prior work, which uses the base-level component to model predict user preferences in various domains (e.g.,~\cite{lex2020modeling,lacic2017beyond,kowald2017temporal,kowald2016influence,Maanen09recommendersystems}), we investigate the \textit{utility of five components}, i.e., (i) base-level, (ii) spreading, (iii) partial matching, (iv) valuation, and (v) noise component, in a next track prediction scenario. These components enable us to investigate the effect of various factors on track relistening behavior, i.e., (i) recency and frequency of prior exposure to tracks, (ii) co-occurrence of tracks, (iii) the similarity between tracks, (iv) familiarity with tracks, and (v) randomness in behavior. To that end, we sequentially simulate track predictions for the next single and remaining track in user sessions considering their previous music consumption on a subset of a publicly available dataset, i.e., the LFM-2b dataset~\cite{melchiorre_ipm_2021}. Our experiments show that recency of prior exposure to tracks is an effective predictor of relistening behavior in terms of R-precision; adding co-occurrence and familiarity in the model further improves the prediction performance.

Summing up, our main contributions are\footnote{
The code is openly available at \url{https://github.com/socialcomplab/recsys21-relistening-actr}
}:
(1) we propose a novel approach of using the declarative memory module within the ACT-R framework to model and predict music relistening behavior, (2) we investigate and discuss the merit of all five declarative memory components of ACT-R in light of what aspects of relistening behavior they let us model.

\section{ACT-R Components}

Our work is based on the declarative memory part of the cognitive ACT-R framework~\cite{anderson2004integrated}. 
In ACT-R, the activation function for a particular chunk $i$ (which in the context of recommender systems represents an item) is defined by the sum of individual components~\cite{bothell2020act} 
given by the equation
\begin{equation}
\label{eq:actr}
    A_i = B_i + S_i + P_i + V_i + \epsilon_i 
\end{equation}
where $B_i$ represents the base-level component, $S_i$ the spreading component, 
$P_i$ the partial matching component, and $V_i$ the valuation component. In addition, $\epsilon_i$ accounts for the noise in the activation. 
The predictions consider all items in a user's music listening history ranked by the activation term $A_i$.
We regularize the results using the softmax function
\begin{equation}
\label{eq:softmax}
    softmax(x_i) = \frac{exp(x_i)}{\sum_{x_j \in I_u} exp(x_j)}
\end{equation}
As a consequence, the sum of all values $x_i$ assigned to candidate items $I_u$, i.e., items in the user history, is $1$. This avoids one component dominating the others, while elevating each component's top items.
In the following, we describe each component in more detail.

\parbold{Base-level component}
The base-level component models item occurrences, considering both their recency and frequency. The component requires a list of tuples $(item, timestamp)$ extracted from the user history as input. Its activation is defined by the base-level learning equation~\cite{kowald2017temporal} given by 
\begin{equation}
\label{eq:bll}
    B_i = softmax\left(\sum^n_{j=1} (t_{ref} - t_{ij})^{-d}\right)
\end{equation}
where $t_{ref}$ represents the reference timestamp, i.e., the timestamp of the prediction, and $t_{ij}$ represents the timestamp of the $j$-th interaction out of $n$ total interactions on item $i$ by the user. The time decay factor $d$ models the forgetting of item interactions and thus balances recency and frequency of occurrences.

The base-level component can approximate, depending on the setting of the parameter $d$, both a user-based most popular (in the case of frequency) or a user-based most recent (in the case of recency) algorithm. More specifically, a base-level component with no decay, i.e., $d=0$, is equivalent to a user-based most popular recommender, while a most recent approach can be approximated arbitrarily close with a very high decay of the base-level component, i.e., $d\rightarrow\infty$.

\parbold{Spreading component}
The spreading component models item co-occurrences and considers $(item, session)$ tuples from the user history as input.
The component \emph{spreads} its activation among items using contextual information, i.e., sessions. Thus, it learns to associate items with each other that appear in the same session, which is modeled using probabilities.
Specifically, the spreading activation~\cite{fum2004memory} is defined by 
\begin{equation}
\label{eq:spreading}
    S_i = softmax\left(\frac{P(i \in C_j)}{P(i)}\right)
\end{equation}
where $P(i)$ represents the probability of item $i$ appearing over all contexts, i.e., past user sessions. $P(i \in C_j)$ represents the probability of item $i$ appearing in context $C_j$. 
We define the last item in the user history as the context item $j$, and the context $C_j$ as all sessions where the context item $j$ appears.
In short, the spreading component predicts items co-occurring often with the most recent item, while also penalizing overall frequent items.
Since frequent occurrences of an item in sessions also lead to high co-occurrences with any other item, this penalty is necessary.

\parbold{Partial matching component}
The partial matching component captures similarities between the items using a list of features. The input tuples are a set (as multiple interactions on the same items do not change the result) and have the form $(item, f_1, f_2, \ldots, f_m)$, where each $f_x$ is an entry of the feature vector of fixed length $m$. These features need to be extracted from the data associated with a given item.  
The activation is then simply defined by a similarity function 
$P_i = softmax\left(sim(i,j)\right)$ 
between the candidate item $i$ and the context item $j$, which again is the last item in the user history. For the similarity function $sim()$, we use the dot product.
This approach is equivalent to a content-based item-to-item recommender, i.e., predicting similar items to those in the target user's profile.

\parbold{Valuation component}
The valuation component was proposed by \cite{juvina2018modeling} and models core affect. 
Instead of assigning affect, e.g., emotions, to items directly, it defines the affective value as a learned parameter by a subject through interactions with stimuli~\cite{juvina2018modeling}. The parameter uses previous interactions and their associated rewards to estimate the valuation of an item.
Specifically, the valuation is defined by the update rule
\begin{equation}
\label{eq:valuation_update}
    V_i(n) = V_i(n-1) + \alpha (R_i(n) - V_i(n-1))
\end{equation}
where $V_i(n)$ specifies the valuation and $R_i(n)$ the associated reward at the $n$-th interaction with a learning rate of~$\alpha$. As a starting point, we set $V_i(0)=0$. As input, the component uses an ordered list (due to the order affecting the update rule) in the form of $(item, reward)$. Thus, we extract the reward directly from the input derived from other properties. 
For music preferences, we can derive the reward from the duration of a listening event, e.g., total time or ratio the track was listened to. %
The activation of the valuation is defined by 
$V_i = softmax\left(V_i(n)\right)$ 
where $V_i(n)$ is calculated by iteratively applying the update rule.

In the simplest case, it applies the familiarity principle, i.e., exposure to an item leads to an increase of preference, which is achieved by setting $R_i(n)=1$ for all interactions. Hence, the component equals an approach that predicts the most popular items of a user.  %
This result could be achieved in a computationally less expensive way by just summing the rewards. However, if negative feedback is incorporated or the rewards differ, then this no longer holds, as the more recent rewards play a bigger role in the valuation term.

\parbold{Noise}
Noise accounts for randomness in behavior in the overall activation function.
Moreover, random predictions are often used as the most basic baseline in recommender systems.
The activation is given by 
$\epsilon_i = softmax\left(rng()\right)$ 
with $rng()$ being a random number generator. On its own, it returns a random item from the list of possible items, i.e., candidate items from the user history. Hence, it takes just a set of $(item)$ tuples as input.

\section{Experimental Setup}

\parbold{Dataset}
Our dataset is a subsample of the LFM-2b dataset\footnote{\url{http://www.cp.jku.at/datasets/LFM-2b/}}~\cite{melchiorre_ipm_2021} from 2019. To reduce the computational demand, we perform stratified sampling\footnote{Based on the distribution, we empirically excluded users with less than $1,000$ and more than $30,000$ listening events and assigned the remaining users to $10$ equal sized bins depending on the number of listening events.} and create a subset of $150$ users with $1,686,296$ listening events (LEs). Thus, our dataset contains $11,242$ LEs per user on average. 
The dataset contains $355,464$ unique tracks, i.e., $5$ LEs on average per track. This small number of average track LEs is explained by the skewed distribution with the most popular track being listened to $1,776$ times, whereas most tracks appear in the long tail, e.g., $158,238$ tracks only appear once in the dataset. This phenomenon outlines the utility of predicting relistening behavior in this dataset as users tend to listen to the same tracks often, i.e., the minority of tracks being responsible for the majority of LEs. 
The prevalence of relistening becomes even more evident considering individual users, who, on average, relisten to previous tracks 66\% of the time and to a single track $111$ times.

We enrich the dataset with session information derived from the timestamps of interactions, and with information of user-generated tags and duration of tracks. Sessions are inferred based on the gaps between LEs, with sessions corresponding to sequences of LEs containing gaps of at most 30 minutes. This is an established approach for session extraction~\cite{eickhoff2014lessons,ortega2010differences} and results in $96,156$ user sessions with on average $18$ LEs per session. The tags and duration of tracks are collected from the Last.fm API\footnote{\url{https://www.last.fm/api/show/track.getTopTags} for tags and \url{https://www.last.fm/api/show/track.getInfo} for duration.}. 
Note that not every track has tags or duration associated with it. As a result, $12,434$ LEs are missing tag information and $108,210$ LEs are missing duration information.

\parbold{Evaluation Protocol}
For each user, we shift a sliding window of one week considering past LEs over all the LEs (including LEs of non-relistened tracks) in the user's listening history, i.e., hop size equals $1$ LE.
Then, we predict the remaining tracks of the current session, i.e., the session from the most recent track. Using this procedure, we simulate $|Q|=811,025$ predictions. 
Considering a fixed time interval instead of the whole user history has the benefit of avoiding scaling issues, while predicting only tracks in the current session provides a realistic prediction scenario. To evaluate the performance, we use the R-precision metric with $R$ being the number of distinct tracks in the remainder of the session given by
\begin{equation}
    R-prec = \frac{1}{|Q|} \sum_{q \in Q} \frac{r}{R}
\end{equation}
with $r$ being the number of relevant tracks evaluated over all generated predictions $q \in Q$. Hence, it captures the ratio of correctly predicted tracks in user sessions, i.e., the precision and recall at $R$.

To contrast this metric, we also evaluate the performance for the next track prediction using the hitrate metric. Given the top predicted track and the next track in the user session, a hit is recognized if the top and next coincide and a miss otherwise. The overall hitrate, i.e., \emph{Next-HR}, is given by averaging the number of observed hits over all possible hits.

\begin{figure*}
\centering
\begin{minipage}[!t]{\textwidth}
  \begin{minipage}[b]{0.62\textwidth}
    \centering
    \includegraphics[width=.85\textwidth,trim={0.2cm 0.2cm 0.2cm 0.2cm},clip]{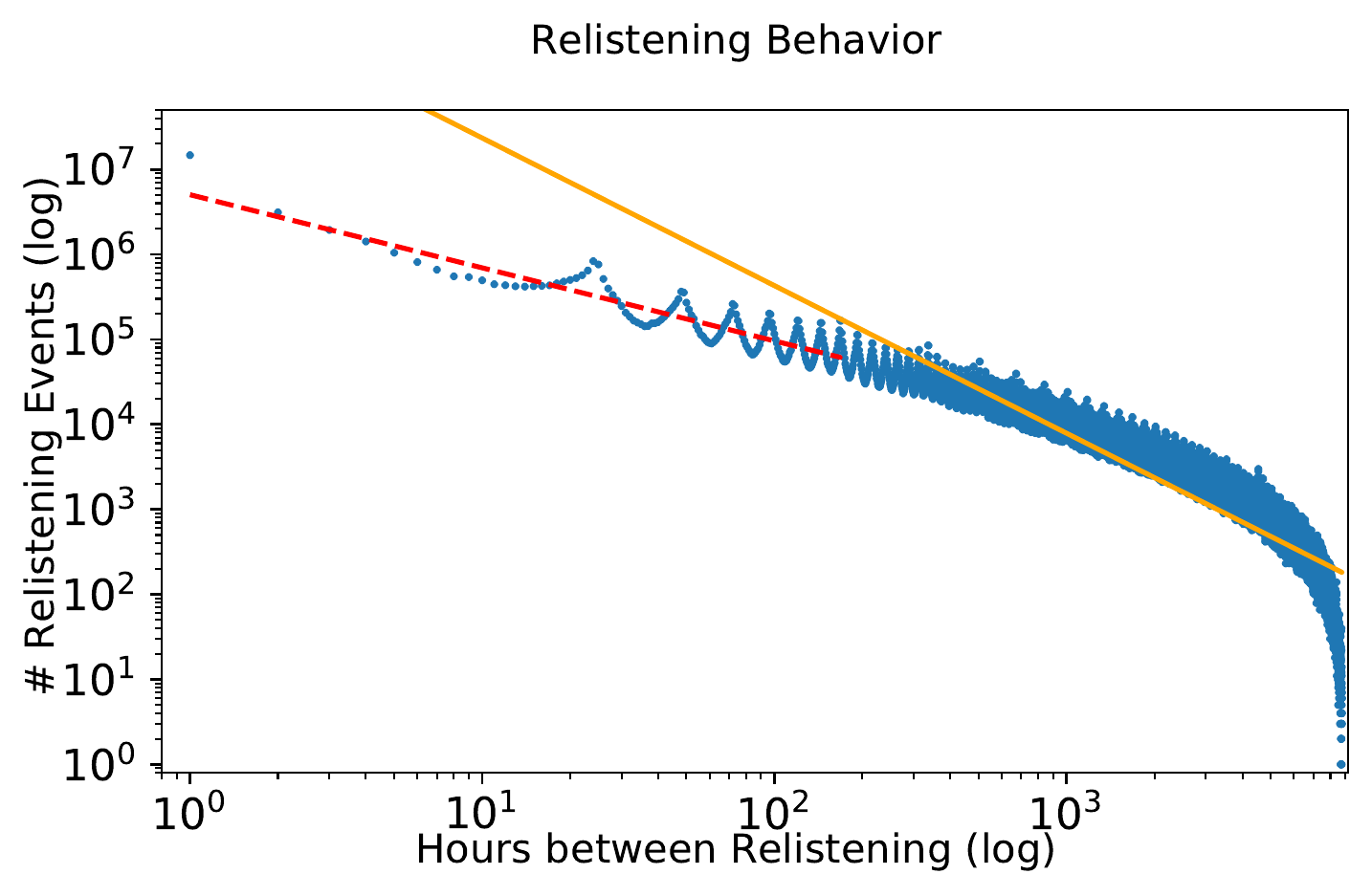}
    \vspace{-0.3cm} %
    \captionof{figure}{Relistening log-log plot of LEs in 2019. Blue dots correspond to value counts of relistening behaviors in hours. Relistening events follow a power law distribution, i.e., have an exponential drop-off in time. The solid orange line (longer, top) is fitted on all LEs in 2019. It provides a good fit of $R^2=0.801$ and has a slope of $-1.737$. We observe an arch at the end of the scale, which we regard as an artifact of filtering by year (i.e., longer relistening behaviors are cut-off). Hence, we also fitted a line considering only shorter relistening behaviors of one week, which is displayed in red (bottom, dashed). This line serves as an estimate for our evaluation protocol and has a slightly better fit of $R^2=0.811$ and a more moderate slope of $-0.860$. %
    The spikes at certain intervals %
    show that the relistening behavior follows some periodicity (e.g., after 1 day). 
    One peak is right at the beginning as people tend to relisten the same track within a short interval (e.g., immediate relistening).}
    \label{fig:relisten}
        \vspace{0.4cm} %
  \end{minipage}
  \vspace{-0.5cm} %
  \hfill
  \begin{minipage}[b]{0.36\textwidth}
    \centering
    \begin{tabular}{lrrrr}
    \toprule
     &  \multicolumn{1}{l}{R-} &  \multicolumn{1}{l}{Next-} \\
    Algorithm & \multicolumn{1}{l}{prec} &  \multicolumn{1}{l}{HR} \\
    \midrule
    TransProb           & \cellcolor{green!5}     .03839 & \cellcolor{green!35}     .15907 \\
    Partial Matching    & \cellcolor{green!5}     .03895 & \cellcolor{green!5}      .01320 \\
    Noise               & \cellcolor{green!5}     .03996 & \cellcolor{green!0}      .00289 \\
    Valuation(discrete) & \cellcolor{green!10}    .04751 & \cellcolor{green!5}      .00533 \\
    Valuation(ratio)    & \cellcolor{green!10}    .05987 & \cellcolor{green!5}      .01042 \\
    Valuation(MP)       & \cellcolor{green!15}    .08436 & \cellcolor{green!5}      .01477 \\
    Spreading           & \cellcolor{green!15}    .09235 & \cellcolor{green!15}     .02117 \\
    Base-level(2019)    & \cellcolor{green!20}    .09903 & \cellcolor{green!20}     .03200 \\
    ACT-R(B,V)          & \cellcolor{green!20}    .10069 & \cellcolor{green!15}     .02416 \\
    MostRecent          & \cellcolor{green!30}    .10167 & \cellcolor{green!25}     .05189 \\
    Base-level(default) & \cellcolor{green!30}    .10380 & \cellcolor{green!15}     .02451 \\
    Base-level(week)    & \cellcolor{green!30}    .10489 & \cellcolor{green!20}     .02883 \\
    ACT-R(S,V)          & \cellcolor{green!35}    .11009 & \cellcolor{green!20}     .02998 \\
    ACT-R(B,S)          & \cellcolor{green!35}    .11042 & \cellcolor{green!20}     .02972 \\
    ACT-R(B,S,V)        & \cellcolor{green!35}    .11119 & \cellcolor{green!20}     .02961 \\
    \bottomrule
    \end{tabular}
    \vspace{0.1cm} %
      \captionof{table}{R-precision of $811,025$ predictions per algorithms on a stratified sample of $150$ users  (\emph{R-prec}) and next track hitrate (\emph{Next-HR}). Base-level is the strongest individual component; combining components (ACT-R with base-level (B), valuation (V), spreading (S)) improves the results. Green shading indicates relative performance.}
      \label{tab:results}
    \end{minipage}
  \end{minipage}
\end{figure*}

\parbold{Algorithm Configurations} 
To configure the decay parameter of the \emph{Base-level} component and in line with previous studies~\cite{lex2020modeling}, we analyze the relistening behavior using a log-log plot as shown in Figure~\ref{fig:relisten}. We observe a good fit for a power law distribution of relistening behavior, i.e., linear fit in the log-log space. Thus, we conjecture that the \emph{Base-level} component, which directly models the decay of such behavior, should already provide good results.
We choose to evaluate the component on three configurations, i.e.,
using the \emph{default} parameter of $d=0.5$ of the ACT-R framework,  
the parameter fitted to the LEs of \emph{2019} with $d=1.737$, and the parameter fitted to the one \emph{week} interval with $d=0.86$.

The \emph{Valuation} component is evaluated deriving the reward from the listening duration compared to the track duration by assigning either an always positive reward equal to the listening \emph{ratio} or mapping it to \emph{discrete} values using $-1$ for $<=33\%$, $1$ for $>=66\%$, and $0$ otherwise. Additionally, we configure it to be equivalent to a user-based most popular (\emph{MP}) approach by setting the reward to $1$ for all LEs. 

For the \emph{Partial Matching} component, we use the user-generated tags for a particular track from which a feature vector is extracted by computing the mean across the tags' word  embedding representation. 
To reduce computational power, we apply Principal Component Analysis %
as a method for dimensionality reduction to the  pre-trained Word2Vec~\cite{mikolov2013efficient} from Google News, whose %
300 dimensions were reduced to 100\footnote{We experimented with other content extraction methods with no avail.}. Finally, the \emph{Spreading} component does not require any parameters. Similarly, there are no parameters for the simulation of \emph{Noise}, as it just predicts random tracks. 

For the combination of the \emph{ACT-R} components, we use the most basic version of each component and denote it as letters, i.e., \emph{B} for base-level with the default value, \emph{S} for spreading, and \emph{V} for valuation with the most popular configuration\footnote{Due to the weak performance, we omitted including the partial matching component and noise.}.
We also included two non-ACT-R approaches in our evaluation for comparison. To contrast the \emph{Spreading} component, we use contiguous sequential pattern mining~\cite{quadrana2018sequence} to estimate the transition probability (\emph{TransProb}) given a reference track. Furthermore, we use an algorithm that predicts the \emph{MostRecent} tracks in the user history.

\section{Experimental Results}

\parbold{Prediction Performance}
Table~\ref{tab:results} reports the prediction performance sorted by R-precision. 
We refer to the algorithms by their names and their configurations (in parentheses). 
\emph{Base-level(week)} is the best performing component in terms of R-precision, which we attribute to the good fit of the power law distribution to the relisting behavior in Figure~\ref{fig:relisten}. \emph{Base-level(default)} is a close second, which demonstrates its applicability even without estimating the decay parameter. In comparison, \emph{Base-level(2019)} performs worse, which hints at the discrepancy between the shorter interval of the evaluation protocol and all LEs in 2019. Notably, \emph{MostRecent} provides a strong baseline, which is also indicated by the left-most data point in Figure~\ref{fig:relisten} that displays an above-average relistening behavior for recent tracks. 
Note that neither a focus on short nor long-term memory has a clear advantage over more balanced approaches that fit the data.

The \emph{Spreading} and \emph{Valuation} components have a lower performance, whereas, \emph{Partial Matching} and \emph{Noise} seem unsuited for the task. 
While \emph{TransProb} is unable to predict the majority of remaining tracks in a session, it has remarkably the best performance on predicting the next single track. Besides, most other algorithms perform similarly well on this task, with the notable exceptions of \emph{MostRecent} performing better and \emph{Noise} performing worst. Here, we also observe that \emph{Partial Matching}, while still under-performing, performs better than random in this regard. We suspect that the low performance of \emph{Partial Matching} is partly explained due to missing tags in the dataset, but also the result of considering tags alone is insufficient for the task, especially when predicting longer sequences. 

Comparing the configuration of \emph{Valuation}, \emph{MP} has the best performance of the three, while \emph{discrete} has the lowest. This observation is more prominent considering the Next-HR metric and can be explained by the strong performance of \emph{MostRecent} and the influence of the last, i.e., the most recent, update step on the final valuation term. Hence, assigning low or negative rewards on recent tracks has a detrimental effect when predicting the next track(s) in a sequence.

Considering the combined models, we observe that three of the four \emph{ACT-R} combinations perform better in terms of R-precision than all individual components. This observation outlines the utility of incorporating multiple cognitive processes for predictions. Notably, \emph{ACT-R(S,V)} outperforms the \emph{Base-level} components without including the component itself. Hence, while \emph{Valuation} and \emph{Spreading} alone only provide mediocre performance, combining them leads to effective predictions. Similar to how \emph{Base-level(2019)} is ill-fitted, \emph{ACT-R(B,V)} leads to a drop that could be the result of a dominance of popular tracks, which are considered by both the \emph{Base-level} and \emph{Valuation} component. In comparison, \emph{ACT-R(B,S)} does not suffer such issue, as the \emph{Spreading} component explicitly penalizes occurrence alone. Finally, \emph{ACT-R(B,S,V)} achieves the highest results as the three components seem to balance each other, i.e., the addition of the \emph{Spreading} component counteracts the dominance of popularity present in \emph{ACT-R(B,V)}.

\parbold{Estimating Component Weights}
In addition to naively combining the components by just adding their activation terms, we explore the applicability of weighted combinations. We evaluate the impact of the three best performing components, i.e., \emph{Base-level(week)}, \emph{Spreading}, and \emph{Valuation(MP)}, on the overall outcome using linear models to weight their activation terms with respect to tracks in the remaining session\footnote{We assign a $1$ to tracks in remaining session and $0$ otherwise. Recall that activation terms lie also in the interval $[0,1]$.} on 10\% of the users. Thus, the altered activation equation becomes $A_i = b\cdot B_i + s\cdot S_i + v\cdot V_i$ with the weights $b$, $s$, and $v$ being estimated by the model. In line with the results in Table~\ref{tab:results}, linear regression models assign the highest weights to \emph{Base-level(week)} and the lowest to \emph{Valuation(MP)}. Considering for instance the linear regression model without fitting an intercept, the learned coefficients are \emph{Base-level(week)} $=0.34$, $Spreading=0.26$, and $Valuation(MP)=0.18$. Evaluating the performance of the weighted combinations does, however, not lead to an increase compared to \emph{ACT-R(B,S,V)}. Moreover, the results are virtually identical to \emph{ACT-R(S,V)}. This phenomenon can be explained when fitting an intercept, as this leads to a slightly negative factor of $Valuation(MP)=-0.05$ without altering the other factors substantially (i.e., $Spreading=0.26$ and \emph{Base-level(week)} $=0.35$). Enforcing a positivity constraint would then nullify the valuation term. This suggests that the valuation term is mostly accounting for the bias, i.e., the fitted intercept, in the model and leads to intuitive results as \emph{Valuation(MP)} scores the most popular tracks highest. Overall, the results indicate that Equation~\ref{eq:actr}, i.e., without weights, is sufficient for predicting music relisting behavior.

\section{Conclusion}

In this paper, we described a psychology-informed approach to model and predict music relistening behavior based on the declarative module of the ACT-R theory. ACT-R consists of several components, i.e., base-level, spreading, partial matching, valuation, and noise, each corresponding to different aspects of memory processes such as recency prior exposure to an item or the familiarity of items. We evaluated our approach on a subset of the LFM-2b dataset by sequentially predicting track relisting from previous listening events of a user. 
Our results show that cognitive aspects of memory retrieval are effective predictors of relistening behavior. Specifically, the recency and frequency of prior exposure as modeled by the base-level component result in the highest predictive performance of individual components.
Besides, we find even more effective combinations by adding spreading and valuation that model co-occurrence and familiarity, respectively. Thus, our work emphasizes the consideration of the whole complexity of user behavior and their underlying cognitive decision-making processes in music predictions.

As a limitation, we recognize that we consider user behavior predictions through the lens of a single yet popular framework from psychology. Also, we evaluate the framework on a single dataset from the music domain on a small number of users (i.e., 150 users with approx. 1.7 Mio. LEs).

In the future, we aim to expand the scope of the experiments by considering more datasets from different domains. Also, we plan to integrate content information such as lyrics via partial matching or other hybrid approaches \cite{deldjoo2021content,deldjoo2020recommender}, rather than only considering interactions. Finally, we see our work not as a replacement, but as a complement for existing approaches, and we will research ways to integrate our findings into classic music recommender systems.

\begin{acks}
We thank Stefan Brandl and Alessandro Melchiorre for their insights and support in data acquisition and preparation. %
\end{acks}

\bibliographystyle{ACM-Reference-Format}
\bibliography{recsys21}
\end{document}